\documentclass[journal,onecolumn]{IEEEtran}
\usepackage{amsmath}
\usepackage{subfigure}
\usepackage{cite}

\usepackage{graphicx}
\usepackage{epstopdf}
\usepackage{multirow}
\usepackage{pgfplots}
\usepackage{siunitx}
\usepackage{nicefrac}
\pgfplotsset{compat=newest}
\usepgfplotslibrary{polar}
\usetikzlibrary{spy}

\newlength\figureheight
\newlength\figurewidth
\newcommand{\figref}{Fig.~\ref}

\newcommand{\secref}{Sec.~\ref}

\usepackage{graphicx}
\usepackage{color}
\usepackage{placeins}
\usepackage{float}
\usepackage{hyperref}
\hypersetup{hidelinks}
\usepackage{tabularx,colortbl}
\usepackage{xcolor}


\usepackage{cite}

\newcommand{\sh}{\textrm{sh}}
\newcommand{\se}{\textrm{se}}

\begin{document}

\title{Exceptional Point Perspective of \\ Periodic Leaky-Wave Antennas}

\author{Amar Al-Bassam$^*$,
        Dirk Heberling${}{^*}{}^\dagger{}$,
        and Christophe Caloz$^\ddagger$ \\
        $^*$Institute of High Frequency Technology, RWTH Aachen University, 52074 Aachen, Germany\\
        ${}{^*}{}^\dagger{}$Fraunhofer Institute for High Frequency Physics and Radar Techniques FHR, 53343 Wachberg, Germany\\
        $^\ddagger$ESAT-TELEMIC, KU Leuven, 3001 Leuven, Belgium
        }

\maketitle
\begin{abstract}
Over the past decade, the issue of gain degradation at broadside in periodic leaky-wave antennas (P-LWAs) has been resolved, using a circuit modeling approach, by introducing proper asymmetry in the unit cell of the antenna structure. This paper provides a more fundamental and insightful perspective of the problem by showing, using a simple coupled-mode analysis, that the optimal level of structural asymmetry corresponds to an exceptional point of the coupling parameter between the two eigenmodes of the P-LWA. This contribution represents a key step towards the development of a full electromagnetic resolution of the broadside issue.
\end{abstract}

\section{Introduction}\label{sec:intro}
Offering the benefits of high directivity and simple scanning, periodic leaky-wave antennas (P-LWAs) represent important, powerful and versatile radiators in the modern microwave, terahertz and optical technology~\cite{Jackson_PIEEE_07_2012}.

Unfortunately, they have been plagued by the issue of gain degradation at broadside~\cite{Jackson_PIEEE_07_2012}. Following a preliminary resolution that consisted of a transmission-line network procedure for unit-cell impedance matching~\cite{Paulotto_TAP_2009}, a more general solution was developed, based on the  fulfillment of the twofold condition of frequency-balancing and $Q$-balancing, with the latter being related to the transverse asymmetry of the unit cell structure~\cite{Otto_TAP_10_2014}. 

The solution introduced in~\cite{Otto_TAP_10_2014} resolves the gain degradation issue, but it is based on circuit modeling that is intricate and that offers little insight into the physics of the problem. This paper presents a more fundamental perspective based on coupled-mode theory and connecting optimal asymmetry with an exceptional point~\cite{Bender_PRL_06_1998}.
\vspace{-0.1cm}
\section{$\mathcal{PT}$-Symmetry and Exceptional Points}\label{sec:EP}
Exceptional points are singularities in the parameter space of a non-Hermitian system where complex eigensolutions coalesce~\cite{Li_PRB_03_2022} as a result of some fundamental symmetry condition. They typically appear in $\mathcal{PT}$-symmetric problems, which involve a specific balance between loss and gain. They may attentively occur in problems involving eigensolutions with distinct levels of dissipation or radiation, which are related to the gain-loss problems by a simple gauge transformation~\cite{Oezdemir_NMAT_08_2019}, as will be shown later to be the case for P-LWAs.

\section{Exceptional Point Derivation}\label{sec:EPD}
\begin{figure}[htbp]
    \centering
    \includegraphics[width=0.6\textwidth]{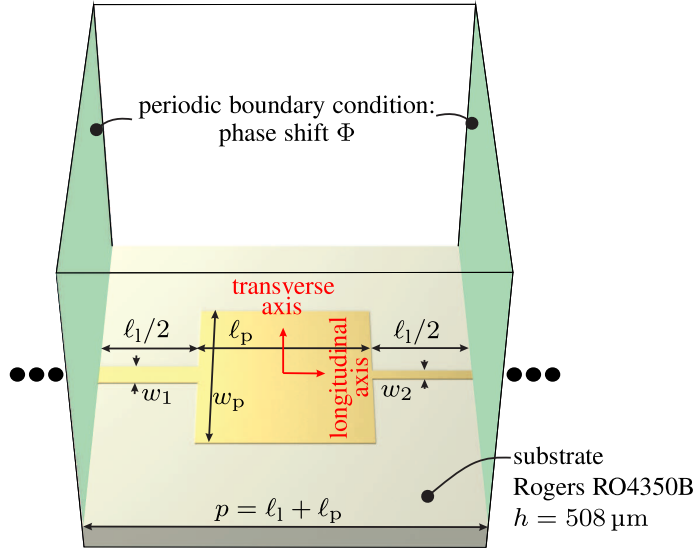}
    \caption{Unit cell of a P-LWA within periodic boundary condition walls (phase shift of $\Phi$ between the two walls). While the structure shown here (and used in the numerical results of \secref{sec:results}) is a series-fed patch structure, the theoretical results of the paper are general and apply to any P-LWA.}
    \label{fig:setup}
\end{figure}

Given their periodicity, P-LWAs are conveniently analyzed in terms of periodic or Floquet-Bloch boundary conditions, as illustrated in \figref{fig:setup}. The enforcement of these conditions with a unit-cell phase shift ($\Phi$) spanning the Brillouin zone provides dispersion diagrams in terms of complex eigenfrequencies ($\Omega$).

If the P-LWA unit cell is symmetric with respect to its transverse axis (see \figref{fig:setup}), it supports two orthogonal eigenmodes, a series mode, with complex eigenfrequency $\Omega_\se$, and a shunt mode, with eigenfrequency $\Omega_\sh$, whose electric fields are  directed along the longitudinal (propagation) and transverse axes of the structure, respectively, and the broadside gain degradation issue ($\Phi=0$) unavoidably occurs~\cite{Otto_TAP_10_2014}. Breaking the transversal asymmetry, couples the series and shunt modes into new, coupled modes, $\psi_1$ and $\psi_2$, which may be described by the coupled-mode equations~\cite{Haus_PIEEE_10_1991}
\begin{subequations}\label{eq:cme}
 \begin{align}   
  \begin{split}     
   \dfrac{d{\psi_1}}{dt} = j\Omega_\se{\psi_1} + j\kappa_{12}{\psi_2},   
  \end{split}\\
  \begin{split}    
   \dfrac{d{\psi_2}}{dt} = j\kappa_{21}{\psi_1} + j\Omega_\sh{\psi_2},
  \end{split} 
 \end{align}  
 \label{eq:coupledEquation} 
\end{subequations}
where $\kappa_{12}$ and $\kappa_{21}$ are the coupling coefficients between the two modes, which are related by reciprocity as $\kappa_{12}=\kappa_{12}^*=\kappa$.

The system~\eqref{eq:cme} admits solutions of the time-harmonic form $\psi_{1,2}\propto\exp(j\Omega t)$. In addition, under the usual frequency-balancing condition~\cite{Jackson_PIEEE_07_2012,Otto_TAP_10_2014}, the uncoupled-mode eigenfrequencies take the form $\Omega_\se=\Omega_0+\Im\{\Omega_\se\}$ and $\Omega_\sh=\Omega_0+\Im\{\Omega_\sh\}$. Substituting all these expressions into~\eqref{eq:coupledEquation} and solving for $\Omega$ yields the coupled eigenfrequency solutions
\begin{subequations}\label{eq:eigsol}
\begin{align}\label{eq:omega_12}
\begin{split}
 \Omega_{1,2} = \Omega_0 &+ j\dfrac{\Im\left\{\Omega_\se\right\} + \Im\left\{\Omega_\sh\right\}}{2} \\
 &\pm \sqrt{\kappa^2 - \left(\dfrac{\Im\left\{\Omega_\se\right\}-\Im\left\{\Omega_\sh\right\}}{2}\right)^2}
 \end{split}
\end{align}
and corresponding coupled $Q$-factors
\begin{align}\label{eq:Q12}
    Q_{1,2}=\frac{\Re\left\{\Omega_{1,2}\right\}}{2\Im\left\{\Omega_{1,2}\right\}}.
\end{align}
\end{subequations}

According to \secref{sec:EP}, the exceptional point of our problem corresponds to the value of $\kappa$ for which the two eigensolutions~\eqref{eq:eigsol} coalesce into a single eigensolution, i.e.,
\begin{align}\label{eq:kappa_opt}
 \kappa_\textrm{opt} = \dfrac{\Im\left\{\Omega_\se\right\}-\Im\left\{\Omega_\sh\right\}}{2},
\end{align}
where we have introduced the subscript ``opt'', for optimal.

Finally, inserting~\eqref{eq:omega_12} with~\eqref{eq:kappa_opt} into~\eqref{eq:Q12} leads to
\begin{align}\label{eq:Q_coupled_opt}
	Q_\textrm{opt} = \dfrac{\Omega_0}{\Im\{\Omega_\se\} + \Im\{\Omega_\sh\}} = \dfrac{2}{\dfrac{1}{Q_\se}+\dfrac{1}{Q_\sh}},
\end{align}
which is exactly the $Q$-balancing formula derived by transmission-line circuit modeling in~\cite{Otto_TAP_10_2014}. This reveals that the solution to the broadside gain degradation issue in P-LWAs coincides with the exceptional point of their level of structural transverse asymmetry.

\section{Validation Results}\label{sec:results}
\begin{figure}[!t]
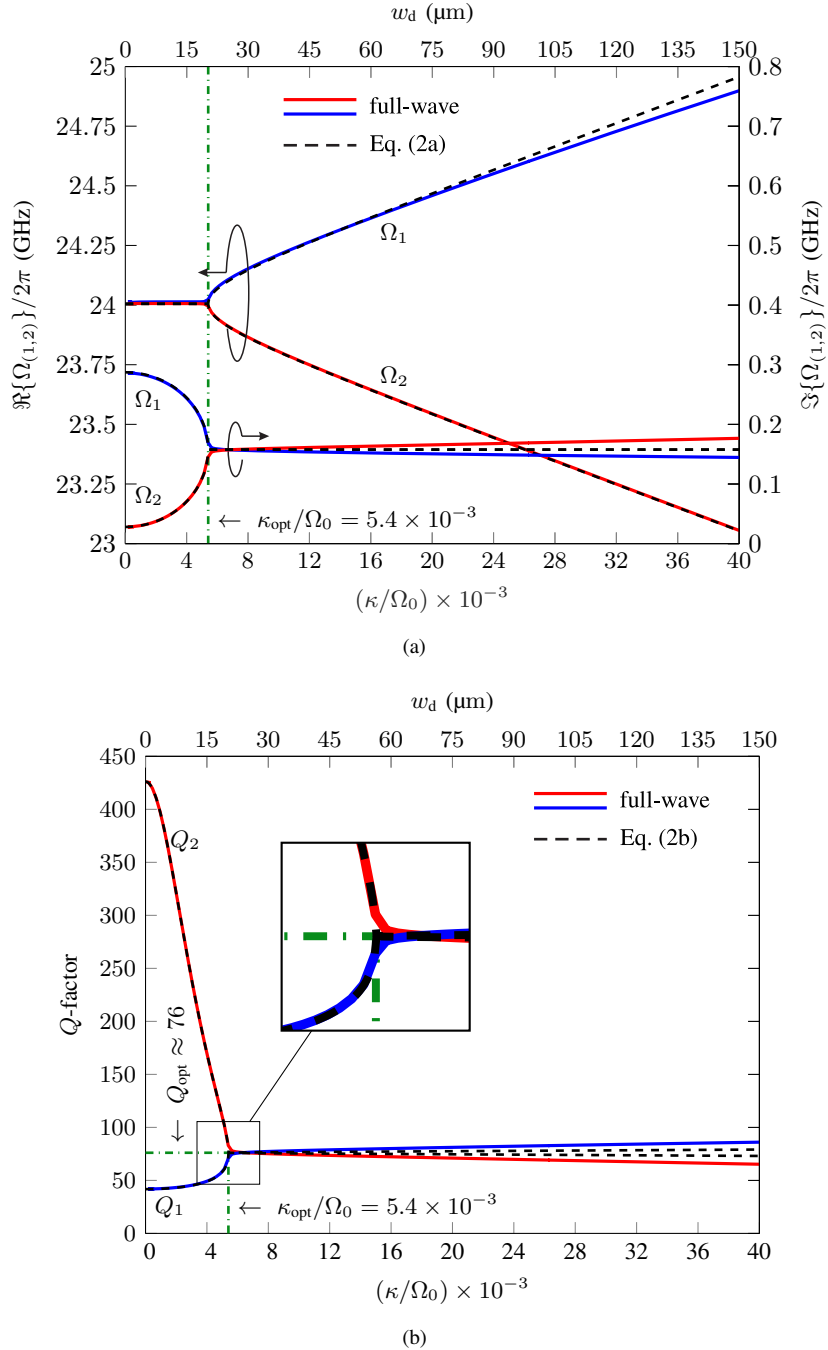

    \centering
    \setlength\figureheight{.35\textwidth}
    \setlength\figurewidth{.45\textwidth}

    \subfigure[]{\input{figs/Omega_CMT_SFP.tikz}\label{fig:Omega}}
    
    \subfigure[]{\input{figs/Q-factor_CMT_SFP.tikz}\label{fig:Q_factor}}
    \caption{Validation of the exceptional point theory established in \secref{sec:EPD} for the series-fed patch P-LWA in \figref{fig:setup}. (a)~Complex eigenfrequencies. (b)~Corresponding $Q$-factors. The optimal coupling factor, given by Eq.~\eqref{eq:kappa_opt}, corresponds indeed to the exceptional point, highlighted by the dash-dotted green line. The unit-cell dimensions are, in mm: $\ell_\textrm{p}=3.05$, $\ell_\textrm{l}=3.5$, $w_\textrm{p}=3.3$ and $w_0=0.3$.}
    \label{fig:results}
\end{figure}

Figure~\ref{fig:results} validates the exceptional point theory of \secref{sec:EPD} for the particular case of the series-fed patch P-LWA in \figref{fig:setup}, with Figs.~\ref{fig:results}(a) and~\ref{fig:results}(a)  plotting the coupled eigenfrequencies and $Q$-factors, respectively.

The curves are obtained as follows. Starting with a symmetric structure, $w_1=w_2=w_0$ (and therefore $\kappa=0$), adjust the lengths of the patch, $\ell_\textrm{p}$, and of the inter-connecting lines, $\ell_\textrm{l}$, in a full-wave eigenfrequency solver (here CST Microwave Studio) so as to frequency-balance the structure, i.e., close up the gap at $\Phi=0$, hence obtaining $\Re\{\Omega_\se\}=\Re\{\Omega_\sh\}=\Omega_0$ and $\Im\{\Omega_{\se,\sh}\}$. Inserting these parameters into~\eqref{eq:eigsol} and plotting the result as a function of $\kappa$ already provides the analytical curves. Then introduce transversal asymmetry in the full-wave eigenfrequency solver by making the widths of the two feeding lines different from each other, specifically setting $w_1=w_0+w_\textrm{d}/2$ and $w_2=w_0-w_\textrm{d}/2$ with increasing $w_\textrm{d}$, with $w_\textrm{d}$ increasing from zero, and plot the corresponding full-wave complex eigenfrequencies and $Q$-factors.

The full-wave results in \figref{fig:results} closely match the analytical predictions\footnote{The slight observed discrepancy, increasing with increasing level of asymmetry, is due to the fact that the frequency-balancing condition, assumed to be fixed in the analytical model, is progressively altered as $w_\textrm{d}$ is increased.}, with the exceptional point appearing at the junction of the two forks formed by the real and imaginary eigenfrequencies and at the junction of the corresponding quality factors, hence validating the central thesis of the paper.
\newpage
\bibliographystyle{IEEEtran}
\vspace{-0.2cm}
\bibliography{IEEEabrv,Bib}
\end{document}